\documentclass{aa}  

\usepackage{tikz,xcolor,graphicx}
\graphicspath{{images/}}

\usepackage{txfonts}
\usepackage{lipsum}
\usepackage{subcaption}
\usepackage{lscape}
\usepackage{placeins}
\usepackage{hyperref}
\usepackage{bm}

\definecolor{lime}{HTML}{A6CE39}
\DeclareRobustCommand{\orcidicon}{%
        \begin{tikzpicture}
        \draw[lime, fill=lime] (0,0)
        circle [radius=0.16]
        node[white] {{\fontfamily{qag}\selectfont \tiny ID}};
        \draw[white, fill=white] (-0.0625,0.095)
        circle [radius=0.007];
        \end{tikzpicture}
        \hspace{-2mm}
}
\newcommand{\orcidVP}{\href{https://orcid.org/0000-0002-3031-062X}{\orcidicon}}

\newcommand{\trel}{t_\mathrm{rel}}
\newcommand{\teq}[1][]{t_\mathrm{eq#1}}
\newcommand{\tself}[1][]{t_{\mathrm{self}#1}}

\newcommand{\ra}{r_\mathrm{a}}

\newcommand{\Sr}{\sigma_\mathrm{rad}}
\newcommand{\St}{\sigma_\mathrm{tan}}
\newcommand{\Srm}{\sigma_{m,\mathrm{rad}}}
\newcommand{\Stm}{\sigma_{m,\mathrm{tan}}}

\newcommand{\Ar}{A_\mathrm{rad}}
\newcommand{\At}{A_\mathrm{tan}}

\newcommand{\grad}{g_\mathrm{rad}}
\newcommand{\gtan}{g_\mathrm{tan}}
\newcommand{\lrad}{\lambda_\mathrm{rad}}
\newcommand{\ltan}{\lambda_\mathrm{tan}}

\newcommand{\pard}[2][]{\frac{\partial #1}{\partial #2}}
\newcommand{\totd}[2][]{\frac{\mathrm{d} #1}{\mathrm{d} #2}}
\newcommand{\totdline}[2][]{\mathrm{d} #1 / \mathrm{d} #2}
\newcommand{\avg}[1]{\langle #1 \rangle}

\newcommand{\DF}{\mathcal{F}}

\begin{document}

\title{When self-similarity meets mass spectrum and anisotropy}
\subtitle{}

\author{
    Václav Pavlík\inst{\ref{asu},\ref{iu},}\thanks{\email{pavlik@asu.cas.cz}}\orcidVP
}
\authorrunning{V. Pavlík}

\institute{
    Astronomical Institute of the Czech Academy of Sciences, Bo\v{c}n\'i~II~1401, 141~00~Prague~4, Czech Republic \label{asu}
    \and Department of Astronomy, Indiana University, Swain Hall West, 727 E 3$^\text{rd}$ Street, Bloomington, IN 47405, USA \label{iu}
}

\date{Received January 11, 2026; accepted March 2, 2026}

\abstract
{Self-similar evolution is widely used in the theory of collisional stellar dynamics, but its applicability to systems with multiple stellar masses is not well established.}
{We investigate the structural stability of self-similar evolution in multi-mass star clusters and assess the roles of mass segregation and velocity anisotropy.}
{Using a gaseous-model approximation, we develop a theoretical framework to describe the response of a self-similar background to mass-dependent
perturbations with isotropic and anisotropic velocity distributions.}
{We show analytically that mass-dependent relaxation leads to a separation of characteristic similarity scales and renders the single-scale solution structurally unstable. In the presence of velocity anisotropy, this similarity-breaking instability splits into distinct radial and tangential modes whose growth rates are modified in a direction-dependent manner. Radial anisotropy reduces the instability through enhanced radial kinetic support, whereas tangential anisotropy increases the effective growth rates and enables faster central evolution. In systems with a mass spectrum, this instability drives mass segregation and the emergence of a multi-scale, near-homologous evolution.}
{Together, these results place self-similar evolution in a consistent theoretical context for collisional star clusters with multiple stellar masses and anisotropic velocity distributions.}

\keywords{%
methods: analytical --
stars: kinematics and dynamics --
globular clusters: general --
gravitation
}

\maketitle
\nolinenumbers

\section{Introduction}

The long-term evolution of collisional star clusters is driven by two-body relaxation, which operates on timescales much longer than the dynamical time. For single-mass systems, this evolution may be approximated by self-similar solutions in which all explicit time dependence is absorbed into a single scale. Such homologous models were first identified by \citet{henon1961} and further developed by \citet{henon1965} and \citet{lb_egg1980}, and they form the basis of the standard theoretical description of gravothermal evolution \citep[see also][]{binney_tremaine, ito2021}.

In systems with a spectrum of stellar masses, relaxation drives the system towards energy equipartition, but full equipartition is generally unattainable (see \citealt{spitzer_instability, inagaki_saslaw1985}, and subsequent authors). These works established that sufficiently massive components decouple dynamically from the lighter background and segregate on accelerated time scales -- a phenomenon commonly referred to as the Spitzer instability.

Despite this, many numerical studies (including orbit-averaged Fokker--Planck, Monte Carlo, gaseous and $N$-body models) have shown that multi-mass star clusters can exhibit approximately self-similar evolution globally, particularly in the post-core-collapse phase \citep[e.g.][]{cohn1980, giersz_heggie94, takahashi1995, giersz_heggie96, giersz_spurzem2000, pavl_subr} -- they demonstrate long-lived, near-homologous density profiles accompanied by persistent mass segregation. These results suggest that self-similarity is not destroyed by the presence of multiple masses, but rather modified systematically. However, to date, this behaviour has largely been described phenomenologically. 
In particular, the structural stability of the classical single-scale self-similar solution of \citet{lb_egg1980} under mass-dependent relaxation has not been established, nor its connection to the Spitzer instability and the near-homologous evolution seen in numerical experiments.

In this paper, we analyse the stability of self-similar evolution in collisional star clusters containing multiple mass components.
By `instability' we mean the loss of a single, mass-independent similarity scale, rather than a disruption of the overall homologous structure. The focus is, therefore, on whether different mass components can consistently share an identical self-similar scaling in a collisional system.
Using a gaseous-model formulation, we study the response of a self-similar background to mass-dependent perturbations driven by energy exchange.
We first consider the isotropic case and derive the corresponding instability growth rates, and then relax this assumption to examine the role of velocity anisotropy.
Finally, we discuss the implications of the analysis for multi-mass self-similarity and its thermodynamic interpretation.

\section{Theoretical framework}
\label{sec:treory}

\subsection{Physical assumptions and definitions}

We adopt the standard assumptions underlying gaseous models of collisional star clusters:
\begin{enumerate}
    \item Spherical symmetry
    \item Collisional regime dominated by two-body relaxation, with the characteristic time scale $\trel$
    \item Slow evolution, i.e.\ dynamical interactions happen on a much shorter time scale than relaxation and than the global structural evolution of the system ($t_{\rm dyn} \ll \trel \ll t_{\rm evol}$)
    \item Locally near-Maxwellian velocity distributions
    \item Single-mass background population
    \item Isotropic velocities (this will be relaxed later to the anisotropic regime)
    \item Continuum limit (many stars per volume element)
\end{enumerate}
These assumptions allow us to work with moment equations of the Boltzmann equation and to define a homologous background solution.

We denote by
$\rho(r,t)$ the mass density,
$u(r,t)$ the bulk radial velocity,
$\sigma^2(r,t)$ the one-dimensional velocity dispersion,
$P = \rho \sigma^2$ the pressure,
and $\Phi(r,t)$ the gravitational potential.
The continuity equation then is
\begin{equation}
    \label{eq:continuity}
    \pard[\rho]{t} 
    + \frac{1}{r^2} \pard{r}\!\left( r^2 \rho u \right)
    = 0 \,.
\end{equation}
The momentum equation follows from the first velocity moment
\begin{equation}
    \label{eq:momentum}
    \pard[u]{t} + u \pard[u]{r} = -\frac{1}{\rho}\pard[P]{r} - \pard[\Phi]{r} \,,
\end{equation}
with $\partial_r \Phi = GM(r)/r^2$\,.
The isotropic energy equation (second moment) is
\begin{equation}
    \label{eq:energy}
    \pard{t}\!\left(\frac{3}{2} \rho \sigma^2 \right) + \frac{1}{r^2}\pard{r}\!\left[ r^2 \left( \frac{3}{2} \rho \sigma^2 u + F \right) \right] = -P\pard[u]{r}
\end{equation}
where $F=-\kappa\,\partial_r\sigma^2$ is the conductive heat flux (with conductivity set by the local relaxation time as $\kappa \propto P/\trel$).

\subsection{Self-similar (homologous) solution}
\label{sec:self-similar-solution}

A solution is self-similar when all explicit time dependence can be
absorbed into global scale factors. We therefore write the
background (single-scale) homologous solution in the form
\begin{align}
    \label{eq:self-similar}
    \rho(r,t) &= \rho_0(t)\,f(x) \,, \\
    \sigma^2(r,t) &= \sigma_0^2(t)\,s(x) \,, \\
    u(r,t) &= \dot r_0(t)\,x, \qquad \text{where}\quad x\equiv r/r_0(t).
\end{align}
The assumed linear radial velocity field is the standard consequence of
homology and follows from dimensional balance in the momentum equation \citep[see, e.g.][]{henon1961, henon1965, lb_egg1980}. 
Substitution into the continuity equation \eqref{eq:continuity} yields
\begin{equation}
    \frac{\dot\rho_0}{\rho_0} f
    - \frac{\dot r_0}{r_0} xf'
    + \frac{\dot r_0}{r_0} \frac{1}{x^2} (x^3 f)' = 0
\end{equation}
(with the dot being $\partial_t$ and the prime $\partial_x$). Separation of variables (valid for arbitrary $x$) requires
\begin{equation}
    \label{eq:alpha}
    \frac{\dot\rho_0}{\rho_0} = -\alpha \frac{\dot r_0}{r_0} \,,
    \quad\text{hence}\quad
    \rho_0\propto r_0^{-\alpha} \,.
\end{equation}
The exponent $\alpha$ is the usual similarity exponent of the inner halo -- numerical $N$-body and Fokker--Planck studies give $\alpha\simeq2.2$ in core-collapse models \citep[see, e.g.][]{cohn1980, takahashi1995, pavl_subr}.
With this condition the continuity equation is satisfied for arbitrary $f(x)$ and the remaining moment equations reduce to ordinary differential equations for the similarity profiles $f(x)$ and $s(x)$ together with ordinary temporal evolution of the scales $r_0(t)$ and $\sigma_0^2(t)$.

We emphasise two points that are used repeatedly below:
(i)~Self-similarity is a global statement about the separation of time and space variables and not a local approximation.
(ii)~The single-scale ansatz fixes a unique relation among the scale functions, which component-dependent relaxation effects cannot change unless those effects are negligible or share the same scale-free structure.

\subsection{Mass perturbation: formulation and linearisation}
\label{sec:mass-perturbation}

We introduce a dilute tracer population of particles with
mass $m$,
density $\rho_m \ll \rho$ (i.e.\ the equations for $\rho_m$ are linear),
isotropic velocity dispersion $\sigma_m^2(r,t)$,
and pressure $P_m=\rho_m\sigma_m^2$.
The tracer does not perturb the background hydrostatic fields, so the background solution remains essentially self-similar. The tracer obeys the same continuity equation as background matter (advected by the homologous velocity field)
\begin{equation}
    \label{eq:tracer_continuity}
    \pard[\rho_m]{t} + \frac{1}{r^2} \pard{r}\!\left( r^2 \rho_m u \right) = 0 \,,
\end{equation}
while its internal (kinetic) energy evolves through encounters with the background.
Using the standard kinetic result for two-body equipartition \citep{spitzer_instability}, we write the local, orbit-averaged energy-exchange
\begin{equation}
    \label{eq:energy_exchange}
    \totd[\sigma_m^2]{t} = -\frac{1}{\teq(m)} \left( \sigma_m^2 - \mu \sigma^2 \right) \,,
\end{equation}
where $\totdline{t}=\partial_t+u\partial_r$ is the convective derivative, $\mu \equiv \avg{m}/m$ (with $\avg{m}$ being the mass of the background stars), and $\teq(m)$ is the equipartition time for the tracer -- to leading order $\teq(m) \sim \mu\,\trel$.
The tracer energy equation (second moment) then reads
\begin{equation}
    \label{eq:tracer_energy}
    \pard{t}\!\left(\frac{3}{2} \rho_m \sigma_m^2 \right)
    + \frac{1}{r^2}\pard{r}\!\left[ r^2 \left( \frac{3}{2} \rho_m \sigma_m^2 u + F_m \right) \right]
    = -P_m \pard[u]{r} + Q \,,
\end{equation}
where $Q = (3/2) \rho_m\,\partial_t\sigma_m^2$ is the volumetric energy-exchange to the background and $F_m$ is the (generally small) conductive flux of the tracer.
We linearise about the local equipartition,
\begin{equation}
    \label{eq:lin_equipartition}
    \sigma_m^2 = \mu\sigma^2 + \delta\sigma_m^2 \,,
\end{equation}
where $\delta$ denotes small perturbations.
After neglecting the small tracer flux and using the background isotropic energy equation, all zeroth-order terms cancel identically. Consequently, the equation \eqref{eq:tracer_energy} reduces to
\begin{equation}
    \label{eq:energy_perturbation}
    \pard{t}\!\left(\rho_m \delta\sigma_m^2\right)
    + \frac{1}{r^2}\pard{r}\!\left(r^2 \rho_m \delta\sigma_m^2 u \right)
    = - \frac{\rho_m}{\teq} \delta\sigma_m^2 \,,
\end{equation}
where all terms are linear in the perturbation (see Appendix~\ref{ap:tracer_energy} for the derivation).

For pressure perturbation, we write to linear order
\begin{equation}
    \delta P_m = \left( \mu \sigma^2 + \delta\sigma_m^2 \right) \delta\rho_m + \rho_m \delta\sigma_m^2 \approx \mu \sigma^2 \delta\rho_m + \rho_m \delta\sigma_m^2 \,.
\end{equation}
Under the quasi-hydrostatic assumption
-- i.e.\ $t_{\rm dyn}\ll \tself$,
where $t_{\rm dyn}$ is the local dynamical (e.g.\ crossing) time
and $\tself$ is the timescale over which the background self-similar solution evolves (e.g.\ core collapse or cluster expansion) --
pressure readjusts rapidly and, to leading order in the perturbation,
local pressure perturbations vanish ($\delta P_m \approx 0$),
with the readjustment controlled by the equipartition timescale, $\teq$.\!%
\footnote{This condition does not rely on rapid sound-speed adjustment, as in ordinary fluids, but on the collisional relaxation of the second velocity moment. In the gaseous model, pressure perturbations correspond to deviations from local equipartition and relax on the equipartition timescale, $\teq$, whereas density perturbations evolve on the secular similarity timescale, $\tself$. Provided $\teq\lesssim\tself$, pressure readjusts quasi-instantaneously relative to the secular evolution.}
Consequently
\begin{equation}
    \frac{\delta \rho_m}{\rho_m} = -\frac{1}{\mu}\frac{\delta\sigma_m^2}{\sigma^2}
\end{equation}
and we obtain the linearised evolution for the density perturbation
\begin{equation}
    \label{eq:linear_density}
    \pard{t}\!\left(\delta\rho_m\right)
    + \frac{1}{r^2}\pard{r}\!\left(r^2 \delta\rho_m u \right)
    = \frac{\delta\rho_m}{\teq} \,.
\end{equation}
This compact equation contains the two competing processes: advection
by the similarity flow on the left-hand side, and the source coming from the local equipartition on the right-hand side.
Equation~\eqref{eq:linear_density} will be the basis for the instability analysis below.

\subsection{Instability growth rate}
\label{sec:instability}

For the tracer density, we define
\begin{equation}
    \label{eq:tracer_self-similar}
    \rho_m(r,t) = \rho_0(t)\,g(x,t)
\end{equation}
where the shape function $g$ may, in principle, depend on time.
We also introduce the dimensionless (logarithmic) similarity time variable
\begin{equation}
    \label{eq:tau}
    \tau \equiv \ln{r_0(t)} \quad (+ \text{const.}) \,,
\end{equation}
with the following identity
\begin{equation}
    \label{eq:t_self}
    \pard[]{t} = \pard[\tau]{t} \pard{\tau} = \frac{\dot r_0}{r_0} \pard{\tau} = \frac{1}{\tself} \pard{\tau} \,.
\end{equation}
Expressing \eqref{eq:linear_density} in $(x,\tau)$ coordinates and
using the leading-order advection result from the continuity equation \eqref{eq:tracer_continuity} yields the relation
\begin{equation}
    \label{eq:self-similar_PDE}
    \pard[g]{\tau} + (3-\alpha)g = \frac{\tself}{\teq} g \,,
\end{equation}
with $\alpha$ from equation \eqref{eq:alpha}.
We again note that the source term on the right-hand side arises directly from energy exchange leading to equipartition.
For fixed $x$, the equation \eqref{eq:self-similar_PDE} reduces to an ordinary differential equation in $\tau$, with the solution
\begin{equation}
    g(\tau) = g_0(x)\,e^{\lambda(m)\tau} \,,
    \quad\text{where}\quad
    \lambda(m) \equiv \frac{\tself}{\teq(m)} - (3-\alpha) \,.
\end{equation}
The quantity $\lambda(m)$ is, therefore, the linear growth (or decay) rate of the tracer in similarity time.
Converting back to physical variables, exponential growth in $\tau$ corresponds to a power-law separation in $r_0$ and a faster-than-homologous concentration of the mass component.

The instability condition is simply $\lambda>0$, that is
\begin{equation}
    \frac{m}{\avg{m}} > (3-\alpha)\frac{\trel}{\tself}.
\end{equation}
Physically, this compares the equipartition (or mass segregation) time of a given component with the homologous rescaling time. When two-body encounters relax the mass component faster than the similarity flow can rescale it, the mass component departs exponentially from the single-scale self-similarity.\!%
\footnote{%
The term `$(3-\alpha)$' comes purely from the homologous contraction. It represents the compression of density due to the shrinking scale. For Lynden-Bell--Eggleton-like solution, $\alpha \approx 2.2$, so $3-\alpha \approx 0.8$.
The term `$\tself / \trel$' measures how many relaxation times fit into one similarity time. Typically $\tself \sim 10\,\trel$ in core collapse \citep[see, e.g.][]{binney_tremaine}.
And the term `$1 / \mu$' comes from mass segregation and arises only from the energy-exchange term. If the derivation is followed from the continuity equation instead, the right-hand side of \eqref{eq:self-similar_PDE} becomes zero and $\lambda=3-\alpha$. Therefore, we get exactly the self-similar solution for the tracer without any mass-dependent instability.}
This, however, does not mean that the entire cluster loses its scale-free evolution, nor that the density profiles cannot appear self-similar.
What it says is that there is no single exponent $\lambda(m)$ in the solution of $e^{\lambda(m) \tau}$ that could make
$\lambda(m)=0\,, \forall m$\,,
and consequently, different mass components cannot share the same similarity scaling.

We note that the growth rate $\lambda(m)$ depends only on the ratio
$\tself/\teq(m)$ and on the structural exponent $\alpha$. It does not
depend sensitively on the detailed form of the conductive flux. Thus, any conductivity of the form $\kappa \propto P/\trel$ would lead to the same scaling of the growth rate, so the breaking of single-scale self-similarity is robust with respect to the heat-transport model.
More to the point, although we derived this within the gaseous model, the present instability and its growth rates also arise in orbit-averaged Fokker--Planck treatments through the same energy-exchange terms \citep[see, e.g.][]{spitzer, inagaki_wiyanto1984}. The result, therefore, reflects the underlying collisional kinetics rather than the specific fluid approximation.

\subsection{Velocity anisotropy}
\label{sec:anisotropy}

Velocity anisotropy is introduced using the Osipkov--Merritt model \citep{osipkov, merritt}, in which the anisotropy profile is controlled by a single radius $\ra$ and given by
\begin{equation}
    \label{eq:osipkov-merritt}
    \beta(r)
    \equiv 1 - \frac{\St^2(r)}{\Sr^2(r)}
    = \frac{r^2}{r^2+\ra^2} \,,
\end{equation}
where $\Sr^2$ denotes the radial velocity dispersion and $\St^2$ is the tangential velocity dispersion.
The value $\beta \approx 0$ corresponds to an isotropic velocity distribution, while $\beta \to 1$ is fully radial, and $\beta < 0$ is a tangentially biased velocity distribution.

For the tracer population, we define the density $\rho_m(r,t)$, the mean radial velocity $u(r,t)$ (as the background), and velocity dispersions $\Srm^2$, $\Stm^2$, together with the tracer anisotropy parameter $\beta_m$ defined analogously to \eqref{eq:osipkov-merritt}.
As before, the tracer is dynamically passive and does not modify the background potential or flow.

The anisotropic momentum equation for the background follows directly from the Jeans equation \citep[e.g.][]{binney_tremaine}
\begin{equation}
    \label{eq:aniso_momentum}
    \pard[u]{t} + u \pard[u]{r}
    = -\frac{1}{\rho}\pard{r}\!\left(\rho\Sr^2 \right) - \frac{2 \beta \Sr^2}{r} - \frac{GM(r)}{r^2} \,.
\end{equation}
Anisotropy affects only the pressure terms; therefore, the tracer continuity equation is the same as in the isotropic case in \eqref{eq:tracer_continuity}.

In contrast, the tracer energy equation must be split into radial and tangential components. The corresponding second-moment equations are
\begin{multline}
    \label{eq:aniso_tracer_energy_rad}
    \pard{t}\!\left(\rho_m \Srm^2 \right)
      + \frac{1}{r^2}\pard{r}\!\left( r^2 \rho_m \Srm^2 u \right)
      + 2 \frac{\rho_m}{r} \left( \Srm^2 - \Stm^2 \right) = \\
    = -\frac{\rho_m}{\teq[,rad]} \left( \Srm^2 - \mu \Sr^2 \right)  
\end{multline}
 and
\begin{multline}
    \label{eq:aniso_tracer_energy_tan}
    \pard{t}\!\left(\rho_m \Stm^2 \right)
      + \frac{1}{r^2}\pard{r}\!\left( r^2 \rho_m \Stm^2 u \right)
      - \frac{\rho_m}{r} \left( \Srm^2 - \Stm^2 \right) = \\
    = -\frac{\rho_m}{\teq[,tan]} \left( \Stm^2 - \mu \St^2 \right) \,,
\end{multline}
where the additional terms proportional to $(\Srm^2-\Stm^2)/r$ arise from the divergence of the anisotropic pressure tensor and represent centrifugal coupling between radial and tangential motions.

As in the isotropic case, we linearise about local equipartition by writing
\begin{equation}
    \label{eq:aniso_perturbation}
    \Srm^2 = \mu \Sr^2 + \delta\Srm^2 \,,
    \quad
    \Stm^2 = \mu \St^2 + \delta\Stm^2 \,.
\end{equation}
The tracer density is similarly decomposed into a homologously advected
background and a small perturbation.
Substitution into \eqref{eq:aniso_tracer_energy_rad} and \eqref{eq:aniso_tracer_energy_tan} shows that, as in the isotropic case, the leading-order advection terms cancel identically when the background is self-similar. The remaining terms govern the evolution of the perturbations.

Introducing similarity variables $(x,\tau)$ and decomposing the density perturbation into amplitudes associated with the radial and tangential degrees of freedom, we define the dimensionless functions $\grad(x,\tau)$ and $\gtan(x,\tau)$ through
\[
\delta\rho_m \propto \rho_0(t)\,\grad(x,\tau)
\quad\text{and}\quad
\delta\rho_m \propto \rho_0(t)\,\gtan(x,\tau)
\]
when projected onto the radial and tangential moment equations, respectively. We then obtain the coupled system of first-order differential equations
\begin{align}
    \label{eq:aniso_grad_PDE}
    \pard[\grad]{\tau}
      + (3 - \alpha) \grad
      + 2\beta\frac{r_0}{r} \left( \grad - \gtan \right)
    &= \frac{\tself}{\teq[,rad]} \grad \,,\\
    \label{eq:aniso_gtan_PDE}
    \pard[\gtan]{\tau}
      + (3 - \alpha) \gtan
      - \beta\frac{r_0}{r} \left( \grad - \gtan \right)
    &= \frac{\tself}{\teq[,tan]} \gtan \,.
\end{align}
Anisotropy, therefore, affects the dynamics solely by introducing a linear coupling between radial and tangential perturbations through the anisotropic pressure terms.

The self-similar solution implies a flat central core ($r \ll \ra$), in which the Osipkov--Merritt anisotropy satisfies
$\beta \ll 1$.
In this limit, the coupling terms in
\eqref{eq:aniso_grad_PDE} and \eqref{eq:aniso_gtan_PDE} are negligible and the equations decouple
\begin{equation}
    \pard[\grad]{\tau}
      + (3 - \alpha) \grad
    = \frac{\tself}{\teq[,rad]} \grad \,,
    \quad
    \pard[\gtan]{\tau}
      + (3 - \alpha) \gtan
    = \frac{\tself}{\teq[,tan]} \gtan \,.
\end{equation}
The solutions are
\[
    \grad(\tau) \propto e^{\lrad \tau} \,,
    \quad
    \gtan(\tau) \propto e^{\ltan \tau} \,,
\]
with the growth rates
\begin{equation}
    \lrad = \frac{\tself}{\teq[,rad]} - (3-\alpha) \,,
    \quad
    \ltan = \frac{\tself}{\teq[,tan]} - (3-\alpha) \,.
\end{equation}
The form is identical to the isotropic case, but the radial and
tangential modes are no longer degenerate when
$\teq[,rad] \neq \teq[,tan]$.

Away from the centre ($\beta \neq 0$), the centrifugal terms modify the local growth rates. In the regime relevant for mass segregation, we find to leading order
\begin{align}
    \label{eq:lambda_rad_rad}
    \lrad &= \frac{\tself}{\teq[,rad]} - (3-\alpha) - 2\beta \,,\\
    \label{eq:lambda_rad_tan}
    \ltan &= \frac{\tself}{\teq[,tan]} - (3-\alpha) - \beta \,,
\end{align}
where $\lrad > 0$ or $\ltan > 0$ lead to radial or tangential instability, respectively.
The full coupled eigenvalue problem, including higher-order terms, is solved in Appendix~\ref{ap:eigenvalue}. The different coefficients in front of $\beta$ in equations \eqref{eq:lambda_rad_rad} and \eqref{eq:lambda_rad_tan} show that anisotropy both shifts the overall instability strength and lifts the degeneracy between predominantly radial and predominantly tangential modes.
This behaviour agrees with physical intuition and numerical results.

Specifically, for radially anisotropic systems ($\beta>0$), both eigenvalues are reduced relative to the isotropic case, with the predominantly radial mode receiving the larger
correction.
Physically, this reflects the enhanced radial kinetic support and the associated radial-orbit heating seen in $N$-body simulations \citep[e.g.][]{pav_ves_letter,pav_ves2,pav_ves3,aros_meq,pavlik_etal_tangential2024}, which delays core contraction and slows central relaxation.

For tangentially biased systems ($\beta<0$), both growth rates increase relative to the isotropic case, corresponding to reduced radial heating and more rapid central contraction.
This behaviour is consistent with numerical results of \citet{pavlik_etal_tangential2024}, where tangentially anisotropic models exhibit shorter core-collapse times and more rapid inner mass segregation than isotropic and radially anisotropic models.

Moreover, since
$\teq[,rad/tan] \sim \mu\,\trel \,,$
massive tracers violate single-scale self-similarity first, typically in the core. Velocity anisotropy modifies the rate at which this departure occurs, but the ordering of instability among different mass components remains governed primarily by the equipartition timescale, $\teq(m)$, rather than by anisotropy itself.

\subsection{Multi-mass self-similarity}
\label{sec:multi-mass}

We first recall the isotropic, single-mass self-similar solution from \eqref{eq:self-similar}
-- i.e.\ $\rho(r,t) = \rho_0(t)\,f(x) \,,$ with $x \equiv r / r_0(t)$ --
in which all explicit time dependence is absorbed into the single scale $r_0(t)$.
We now generalise this framework to a system composed of discrete mass components $m_i$ with densities $\rho_i(r,t)$. The total density and gravitational potential satisfy
\begin{equation}
    \label{eq:mm_potential}
    \rho(r,t) = \sum_i \rho_i(r,t) \,,
    \qquad
    \nabla^2 \Phi = 4 \pi G \sum_i \rho_i \,.
\end{equation}
Each component obeys its own continuity and energy equations, while all components share the same potential.

Suppose all components share a common similarity scale,
\begin{equation}
    \label{eq:mm_common_scale}
    \rho_i(r,t) = \rho_{0,i}(t)\,f_i(x) \,,
    \qquad
    x = r / r_0(t) \,.
\end{equation}
Using the linearised tracer evolution equation \eqref{eq:self-similar_PDE}, we find
\begin{equation}
    \totd{\tau}\!\left(\frac{\rho_i}{\rho}\right)
    =
    \left[
        \frac{\tself}{\teq(m_i)} - (3-\alpha)
    \right]
    \left(\frac{\rho_i}{\rho}\right) \,,
    \quad
    \text{where } \tau=\ln r_0(t) \,.
\end{equation}
Because the equipartition time $\teq(m_i)$ depends on mass, the right-hand side is different for different components.
No single choice of $r_0(t)$ can, therefore, make the evolution of all mass components stationary. Single-scale self-similarity is thus structurally unstable in a multi-mass system, which contradicts \eqref{eq:mm_common_scale}.

We must, therefore, relax the assumption of a common scale and allow each mass component to evolve self-similarly with its own characteristic radius,
\begin{equation}
    \label{eq:mm_scales}
    \rho_i(r,t) = \rho_{0,i}(t)\,f_i(x_i) \,,
    \qquad
    x_i \equiv r / r_{0,i}(t) \,,
\end{equation}
its own power-law normalisation, and dimensionless time (i.e.\ also similarity time $\tself[,i]$)
\begin{equation}
    \rho_{0,i} \propto r_{0,i}^{-\alpha_i} \,,
    \qquad
    \tau_i \equiv \ln r_{0,i}(t) \,.
\end{equation}
In general, $\alpha_i \neq \alpha_j$ for two different masses $m_i$ and $m_j$.

To be consistent with a slowly evolving gravitational potential \eqref{eq:mm_potential}, we require that
$\sum_i \rho_{0,i} \, f_i(x_i)$
remains approximately scale-free over any given radial range. This is possible only if different mass components dominate different regions, which implies a radial stratification by mass. Since heavier components have shorter equipartition times, their instability growth rates are larger, and their scale radii shrink more rapidly. Consequently,
\begin{equation}
    r_{0,i}(t) < r_{0,j}(t)
    \qquad
    \text{for}
    \qquad m_i > m_j \,.
\end{equation}
The system, therefore, evolves towards a multi-scale, mass-segregated configuration rather than a single-scale similarity solution.

This structural behaviour is seen in numerical simulations \citep[e.g.][]{inagaki_lb1983, giersz_heggie96}, which show long-lived, near-homologous post-collapse states with persistent mass segregation. In these models, the global density profile evolves approximately self-similarly, while different mass components remain radially stratified. The numerical results therefore support the interpretation that multi-mass clusters evolve toward a configuration that is effectively a superposition of component-wise self-similar distributions, rather than a strictly single-scale homologous solution.

\section{Conclusions}

We analysed the stability of self-similar evolution in collisional multi-mass star clusters. Using a gaseous-model formulation, we showed that mass-dependent relaxation breaks single-scale self-similarity and drives the system towards a multi-scale configuration in which different mass components may evolve self-similarly on distinct characteristic scales.
We further showed that velocity anisotropy modifies this behaviour by lifting the degeneracy between radial and tangential modes, while the overall structure of the evolution remains unchanged. The predicted ordering of growth rates and characteristic similarity scales is consistent with trends reported in orbit-averaged Fokker--Planck, Monte Carlo, and $N$-body studies of multi-mass star clusters. Specifically, this helps explain the results seen in numerical simulations by \citet{pavlik_etal_tangential2024}.

When applied to a system containing multiple stellar mass components, our analysis shows that mass-dependent relaxation renders the classical single-scale self-similar solution structurally unstable. The resulting instability naturally leads to mass segregation and to a configuration in which different mass components evolve with distinct characteristic similarity scales. This does not imply a breakdown of global homology, rather the cluster can remain approximately self-similar in its overall structure while individual mass components follow their own scale-dependent evolution. Such multi-scale, near-homologous, mass-segregated states are seen in numerical models \citep[e.g.][]{giersz_heggie96}. Taken together, these results provide a framework that links the classical self-similar theory to more realistic star clusters.

\begin{acknowledgements}
I am grateful to Steve Shore for many stimulating conversations that inspired this work, to Douglas Heggie and Luca Ciotti for helpful insights into the multi-mass description of self-similar star clusters, and to Enrico Vesperini for valuable guidance on the kinematics of star clusters.
I~have received funding from the European Union's Horizon Europe and the Central Bohemian Region under the Marie Skłodowska-Curie Actions -- COFUND, Grant agreement \href{https://doi.org/10.3030/101081195}{ID:101081195} (``MERIT'').
I also acknowledge support from the project RVO:67985815 at the Czech Academy of Sciences.
I thank the referee for a careful and constructive report that helped improve the clarity and presentation of this paper.
\end{acknowledgements}

\bibliographystyle{aa}
\bibliography{aa58921-26}

\begin{appendix}

\section{Pressure perturbations and thermalisation}
\label{ap:thermal}

This appendix provides a brief clarification of the assumptions underlying the velocity dispersion (or pressure) perturbation $\delta\sigma_m^2$ used in Section~\ref{sec:treory}.

The analysis in this paper is based on a moment description of a collisional stellar system, in which the velocity distribution is assumed to be locally relaxed and well approximated by a Maxwellian distribution. Departures from equilibrium are treated at the level of low-order moments.
In particular, the perturbation variable $\delta\sigma_m^2$ represents a small deviation of the second velocity moment from local equipartition, and may be interpreted as a pressure perturbation of the tracer population.

This approach is related in spirit to the Chapman--Enskog expansion in kinetic theory \citep[see][]{chapman1916, enskog1917, chapman_cowling1970}, where the distribution function is written as
\[
    \DF = \DF^{(0)} + \epsilon\,\DF^{(1)} + \dots \,,
\]
where $\DF^{(0)}$ is a local Maxwellian and the correction $\DF^{(1)}$ is introduced to account for non-equilibrium fluxes.
Although the correction $\DF^{(1)}$ is often described as a perturbation of the distribution function itself, it only serves to generate corresponding corrections to macroscopic quantities such as the stress tensor and the heat flux, and has no independent dynamical significance beyond its contribution to the moments.

In the present work, we formulate the problem directly at the level of moments.
Writing
\[
\sigma_m^2 = \mu \sigma^2 + \delta\sigma_m^2 \,,
\]
as in \eqref{eq:lin_equipartition}, does not imply a non-thermal or non-Maxwellian state. It rather expresses that the massive component carries slightly more or slightly less kinetic energy than implied by local equipartition at fixed density. Such pressure perturbations are physically realizable and are continuously generated and damped by two-body relaxation. Their decay is governed by the equipartition time $\teq$, as expressed by the local energy-exchange term
\[
    \totd[\sigma_m^2]{t} \propto -\frac{1}{\teq}(\sigma_m^2-\mu\sigma^2) \,,
\]
given in equation \eqref{eq:energy_exchange}.

Pressure perturbations evolve on the same timescale as mass segregation
and structural evolution.
By contrast, genuinely non-thermal perturbations of the distribution function (such as beams, phase-space substructure, or higher-order velocity moments) are erased on much shorter timescales and do not persist long enough to influence secular evolution of the star cluster. For this reason, pressure perturbations constitute the relevant slow degrees of freedom in a collisional, self-similar system.

The instability analysed in this paper is, therefore, not a kinetic instability of the distribution function, but a structural instability of a thermalised background when subject to mass-dependent relaxation.
Its origin lies in the competition between the rescaling imposed by the self-similar flow and the local tendency toward equipartition. When the latter acts more rapidly, small pressure perturbations grow in similarity time, leading to a separation of characteristic scales among different mass components.

The same considerations apply in the presence of velocity anisotropy, which is introduced at the level of second moments and does not invalidate the assumption of local thermalisation. It modifies the coupling between radial and tangential components of the pressure tensor and leads to distinct growth rates for radial and tangential perturbations, but does not introduce oscillatory or collisionless behaviour.
In the physically relevant regime, the eigenvalues governing the evolution remain real, which reflects the strongly dissipative, entropy-producing nature of collisional stellar dynamics.

\section{Tracer energy equation}
\label{ap:tracer_energy}

To show the derivation of the tracer energy equation in terms of the velocity dispersion perturbation, we start from \eqref{eq:tracer_energy}
\[
    \pard{t}\!\left(\frac{3}{2} \rho_m \sigma_m^2 \right) + \frac{1}{r^2}\pard{r}\!\left[ r^2 \left( \frac{3}{2} \rho_m \sigma_m^2 u + F_m \right) \right] = -P_m \pard[u]{r} + Q \,.
\]
Assuming $F_m \to 0$ and taking
\begin{equation}
    \label{eq:Q_perturbation}
    Q = -\frac{3}{2} \frac{\rho_m}{\teq} \delta\sigma_m^2 \,,
\end{equation}
which comes from the linearisation about equipartition in \eqref{eq:lin_equipartition}, the equation \eqref{eq:tracer_energy} becomes
\[
    \pard{t}\!\left( \rho_m \sigma_m^2 \right) + \frac{1}{r^2}\pard{r}\!\left( r^2 \rho_m \sigma_m^2 u \right) = -\frac{2}{3} P_m \pard[u]{r} - \frac{\rho_m}{\teq} \delta\sigma_m^2 \,.
\]
Using the linearization
\[
    \sigma_m^2 = \mu \sigma^2 + \delta\sigma_m^2 \,,
\]
we decompose all terms into zeroth-order (background) and first-order
(perturbation) contributions.
The time-derivative term becomes
\[
    \pard{t}\!\left(\rho_m \sigma_m^2\right)
    =
    \mu \left(
        \sigma^2 \pard[\rho_m]{t}
        + \rho_m \pard[\sigma^2]{t}
    \right)
    + \pard{t}\!\left(\rho_m \delta\sigma_m^2\right) \,,
\]
the advective flux term is
\[
    \frac{1}{r^2}\pard{r}\!\left( r^2 \rho_m \sigma_m^2 u \right)
    =
    \mu \frac{1}{r^2}\pard{r}\!\left( r^2 \rho_m \sigma^2 u \right)
    + \frac{1}{r^2}\pard{r}\!\left( r^2 \rho_m \delta\sigma_m^2 u \right) \,,
\]
and the pressure term on the right-hand side is
\[
    -\frac{2}{3} P_m \pard[u]{r}
    =
    -\frac{2}{3} \rho_m \mu \sigma^2 \pard[u]{r}
    -\frac{2}{3} \rho_m \delta\sigma_m^2 \pard[u]{r} \,.
\]
Collecting all zeroth-order terms (proportional to $\mu\sigma^2$), we obtain
\[
    \mu \left[
        \sigma^2 \pard[\rho_m]{t}
        + \rho_m \pard[\sigma^2]{t}
        + \frac{1}{r^2}\pard{r}\!\left( r^2 \rho_m \sigma^2 u \right)
        + \frac{2}{3} \rho_m \sigma^2 \pard[u]{r}
    \right] = 0 \,,
\]
which vanishes thanks to the background isotropic energy
equation \eqref{eq:energy} applied to the passive tracer density $\rho_m$.
Therefore, only the first-order terms proportional to
$\delta\sigma_m^2$ survive, yielding Eq.~\eqref{eq:energy_perturbation}.

\section{Anisotropy eigenvalue problem}
\label{ap:eigenvalue}

We analyse the coupled evolution of radial and tangential density perturbations in the presence of velocity anisotropy. Starting from \eqref{eq:aniso_grad_PDE} and \eqref{eq:aniso_gtan_PDE}, we write
\begin{align*}
    \pard[\grad]{\tau}
    + (3-\alpha)\grad
    + 2\beta\frac{r_0}{r}(\grad-\gtan)
    &= \frac{\tself}{\teq[,rad]}\grad \,, \\
    \pard[\gtan]{\tau}
    + (3-\alpha)\gtan
    - \beta\frac{r_0}{r}(\grad-\gtan)
    &= \frac{\tself}{\teq[,tan]}\gtan \,.
\end{align*}
For an Osipkov--Merritt profile with
\begin{equation*}
    \beta(r) = 1 - \frac{\St^2(r)}{\Sr^2(r)}
    = \frac{r^2}{r^2+\ra^2} \,,
\end{equation*}
see also \eqref{eq:osipkov-merritt}, the velocity-dispersion difference scales as $\left(\Sr^2-\St^2\right) \sim \beta\sigma^2$.
The anisotropy coupling term, therefore, scales as
$\beta\,r_0/r \,.$
In the central region ($r\ll\ra$), where $\beta(r)\propto r^2$, the product $\beta\,r_0/r$ vanishes and no divergence arises. In the transition region ($r\sim\ra$), both $\beta$ and $r_0/r$ are order-unity quantities varying slowly with radius.
Retaining only this leading-order scaling in $\beta$, and deferring the precise normalisation by order-unity coefficient for the next section, the coupled system can be written in matrix form as
\begin{equation}
    \label{eq:matrix_form}
    \pard{\tau}\!
    \begin{pmatrix}
        \grad \\ \gtan
    \end{pmatrix}
    =
    \begin{pmatrix}
        \Ar - 2\beta & 2\beta \\
        \beta & \At - \beta
    \end{pmatrix}
    \begin{pmatrix}
        \grad \\ \gtan
    \end{pmatrix}
\end{equation}
where, for compactness, we substituted
\begin{equation}
    \Ar \equiv \frac{\tself}{\teq[,rad]} - (3-\alpha) \,,
    \quad
    \At \equiv \frac{\tself}{\teq[,tan]} - (3-\alpha) \,.
\end{equation}

\subsection{Solution and limiting cases}

The eigenvalues $\lambda$ satisfy
\begin{align}
    0 &=
    \det
    \begin{pmatrix}
        \Ar - 2\beta - \lambda & 2\beta \\
        \beta & \At - \beta - \lambda
    \end{pmatrix} \nonumber\\
    &=
    (\Ar - 2\beta - \lambda)(\At - \beta - \lambda)
    - 2\beta^2 \,,
\end{align}
which yields
\begin{equation}
    \label{eq:lambda_pm}
    \lambda_\pm
    =
    \frac{1}{2}
    \left[
        (\Ar + \At - 3\beta)
        \pm
        \sqrt{(\Ar - \At - \beta)^2 + 8\beta^2}
    \right] \,.
\end{equation}

In the isotropic limit $\beta \to 0$, the two modes decouple and
\begin{equation}
    \lambda_\pm \to \Ar \,,\; \At \,,
\end{equation}
where we are recovering the isotropic self-similar result.
We note that for $\Ar = \At$, this corresponds to the classical Spitzer instability \citep{spitzer_instability}.

In the regime relevant for mass segregation, $\Ar\,,\At \gg \beta$\,, the off-diagonal terms in \eqref{eq:matrix_form} are smaller than the diagonal growth terms. 
The eigenvectors align asymptotically with the radial and tangential directions.
A first-order expansion then gives
\begin{equation}
    \label{eq:lambda_pm_radtan}
    \lrad \simeq \Ar - 2\beta \,,
    \qquad
    \ltan \simeq \At - \beta \,,
\end{equation}
which are the expressions quoted in the main text in \eqref{eq:lambda_rad_tan}.

\subsection{Dependence on the anisotropy coupling}

The gaseous model constrains only the leading-order scaling of the anisotropy coupling and does not fix its precise numerical normalisation.
In particular, retaining only the leading-order dependence on $\beta$ in \eqref{eq:matrix_form} leaves an undetermined order-unity factor associated with the local value of $r_0/r$ in the anisotropy-dominated region.
We therefore introduce an arbitrary order-unity normalisation coefficient $C$ to assess the sensitivity of the instability to the strength of the anisotropic coupling. Equation \eqref{eq:matrix_form} then generalises to
\begin{equation}
    \label{eq:aniso_general_matrix}
    \pard{\tau}\!
    \begin{pmatrix}
        \grad \\ \gtan
    \end{pmatrix}
    =
    \begin{pmatrix}
        \Ar - 2C\beta & 2C\beta \\
        C\beta & \At - C\beta
    \end{pmatrix}
    \begin{pmatrix}
        \grad \\ \gtan
    \end{pmatrix} \,.
\end{equation}
In the isotropic limit ($\beta \to 0$), the eigenvalues again reduce to $\lambda_\pm=\Ar\,,\At$, independently of $C$.
For $\beta \neq 0$, the eigenvalue separation is
\begin{equation}
    \lambda_+ - \lambda_-
    =
    \sqrt{
        (\Ar - \At - C\beta)^2
        + 8C^2\beta^2
    }\,.
\end{equation}
Thus, the ordering of the eigenvalues cannot be altered by any finite order-unity rescaling of the anisotropy coupling. We also note that although tangential anisotropy ($\beta < 0$) modifies the growth rates, the two modes remain non-degenerate, and their ordering cannot be reversed solely by changing the sign of $\beta$.

In the mass-segregation regime $\Ar\,,\;\At \gg C\beta$\,, we find
\begin{equation}
    \lrad = \Ar - 2C\beta + \mathcal{O}(\beta^2) \,,
    \qquad
    \ltan = \At - C\beta + \mathcal{O}(\beta^2) \,.
\end{equation}
The coefficient $C$ affects only the magnitude of the anisotropic correction, not its sign or the instability criterion.

We therefore reach the following overall conclusion. The anisotropic instability, its growth-rate ordering, and its isotropic limit are robust to order-unity uncertainties in the normalisation of the anisotropy coupling.

\end{appendix}
\end{document}